# Physics-Informed Neural Network Based Digital Image Correlation Method

Boda Li[a], Shichao Zhou[a], Qinwei Ma[b], Shaopeng Ma[a, *]
([a] School of Ocean and Civil Engineering, Shanghai Jiao Tong University, Shanghai 200240, China; [b] School of Aerospace Engineering, Beijing Institute of Technology, Beijing 100081, China)
*Corresponding author：mashaopeng@sjtu.edu.cn

## Abstract

Digital Image Correlation (DIC), a widely used full-field deformation measurement technique in experimental mechanics, traditionally employs subset matching to ascertain the displacement field. In scenarios characterized by non-uniform deformation, however, challenges arise in selecting such optimal parameters as shape functions and subset size. Recent developments in deep learning-based DIC, including supervised and unsupervised learning approaches, have leveraged neural networks to establish a direct and complex mapping between speckle images and deformation fields. The inherent robustness of neural networks in nonlinear fitting enables these approaches to yield precise measurements in non-uniform fields, thereby circumventing the need for manual parameter tuning. Nonetheless, as the existing deep learning-based DIC frameworks take the speckle image as the input and the displacement field as the output, its prerequisite is to extract the complex features of the speckle image before fitting the displacement field, which demands an elaborate network architecture, yet provides no guarantee of solution accuracy. This paper proposes a method referred to as PINN-DIC, a novel DIC method based on Physics-Informed Neural Networks (PINNs), which utilizes a simple fully connected neural network that takes the coordinate domain as input and outputs the displacement field. The photometric consistency assumption, the DIC governing equation, is integrated into its loss function and the displacement field can be directly extracted from reference and deformed speckle images by iteratively optimizing the network parameters. The principles and solution process of PINN-DIC are introduced, and the method is evaluated and validated through both simulated and real experiments. PINN-DIC inherits the conventional advantages of deep learning-based DIC in the measurement of non-uniform fields, i.e., its capacity to accurately resolve non-uniform displacement fields without pre-defined parameters. Additionally, PINN-DIC presents three distinct advantages attributable to its network architecture: 1) it achieves enhanced precision with a comparatively lighter network by directly fitting the displacement field from coordinates rather than processing the speckle images; 2) it effectively addresses displacement fields in speckle images with irregular boundaries with minor adjustments to the input parameters; and 3) it can be easily integrated with other neural network-based mechanical analysis methods, enabling deeper integration and analysis of DIC measurement results.

# 1 Introduction

Digital Image Correlation (DIC) is a classical method for full-field deformation measurement in experimental mechanics [1], widely applied in both scientific research and engineering fields. The determination of the displacement field in DIC is formulated as an inverse problem: with the grayscale fields of two speckle images captured before and after deformation (respectively identified as the reference and deformed images), the displacement field is resolved by imposing the photometric consistency assumption, i.e., the grayscale at a given point remains unchanged before and after deformation. The strain field is subsequently derived through the numerical differentiation of the displacement field. Given that the number of displacement vector components (two) at each point in the image exceeds the number of grayscale scalar components (one), DIC is an ill-posed problem and often requires additional constraints to solve for the displacement vector at each point. The subset-based traditional DIC method (Subset-DIC [2, 3]), delineates the image into a series of independent subsets and solve the displacement within each subset. By incorporating rigid body or linear displacement assumptions to a subset (based on the assumption that the subset is small and then the full-field deformation is simple), the subset's unknown displacement variables are substantially reducing to 2 or 6, a figure markedly less than the subset's points, effectively reducing the ill-posed nature of the problem and rendered it solvable. Similarly, a mesh-based DIC method (mesh-DIC [4]) employs a similar approach by meshing the image and assuming the displacement within the mesh elements follows a linear or quadratic variation pattern, thereby significantly reduces the number of unknown displacement variables and makes the problem solvable. The efficacy of both Subset-DIC and mesh-DIC is profoundly contingent upon the displacement representation assumptions within subsets or mesh elements, known as shape functions, and the size of these subsets or mesh elements [4-6]. However, the selection of appropriate shape functions and element sizes of subset or mesh depends on multiple factors, such as the degree and scale of deformation non-uniformity [7-10], image noise levels [11] and speckle size [12, 13]. A plethora of studies have explored these influencing factors, but a conclusive and universal theoretical framework has yet to emerge [14-16]. Consequently, high-quality DIC analysis results in practical measurements often rely on the accumulated experience of the measurement personnel. After the experiment is completed, the determination of such parameters as subset shape functions and sizes is often predicated on an estimation of specimen's deformation, speckle characteristics, noise characteristics, and other factors. This dependence on experiential knowledge for the selection of parameters substantially impedes the broader application of DIC methodologies within the engineering sector, particularly the measurement of complex non-uniform deformation fields [8, 17].

Researchers are progressively pivoting towards deep learning techniques to enhance Digital Image Correlation (DIC), with efforts focused on both supervised and unsupervised learning methods [18-26]. In supervised learning for DIC, Boukhtache et al. [18] employed a U-net network combining downsampling and upsampling convolutional neural networks to solve the displacement field from speckle images. Yang et al. [19] used two separate U-net networks to solve the displacement and strain fields. Cheng et al. [20] added mechanical constraints to the U-net network, incorporating both displacement and strain terms into the loss function to improve the accuracy of the displacement field solution. Duan et al. [21] extended this method to Digital Volume Correlation (DVC) by employing three convolutional sub-networks for overall sub-voxel searching, sub-voxel registration, and displacement field denoising in 3D deformation analysis [22]. The fundamental idea that underpins supervised learning DIC is to construct a direct mapping between pairs of speckle images (reference and deformed images) and displacement fields using neural networks. This process involves the training of the network and optimization of its parameters using an extensive dataset of known mappings generated from simulated speckle images and their corresponding standard displacement fields. Due to the powerful nonlinear adaptive fitting capability of neural networks, deep learning DIC are capable of markedly surmounting the challenges in displacement representation and subset/mesh element size selection found in Subset-DIC approaches, particularly in measuring non-uniform fields [18-20]. However, the training of the network is fundamentally a data-driven endeavor, necessitating an extensive collection of annotated data, which in the context of DIC, refers to pairs of speckle images (reference and deformed images) and their corresponding deformation fields (displacement or/and strain fields). In practical operations, the initial labeling for supervised DIC is predominantly derived from simulated data [18, 19] due to the challenging procurement of ideal standard deformation fields from real experimental data and the request for the accuracy of labeled data during the training process. This training method results in limited generalization ability of supervised learning DIC algorithms, meaning they fail to achieve high-precision results when processing actual experimental speckle images, especially in the presence of noise. To address this issue, Pan et al. [23] introduced concepts from optical flow methods to enhance the accuracy and generalization performance of supervised learning DIC with improved network structures and training datasets. To thoroughly tackle the problems of low generalization ability and accuracy in supervised learning DIC caused by its reliance on simulated speckle images, Cheng et al. [24], Wang et al. [25], and Zhu et al. [26] independently developed unsupervised learning DIC methods, which retains the employment of neural networks to represent displacement fields while amending the formulation of the loss function. Instead of comparing the predicted displacement field with a standard displacement field, it compares the predicted deformed image generated from the reference image based on photometric consistency with the actual deformed image. When the physical governing equation, i.e., photometric consistency, is incorporated between speckle images, unsupervised learning DIC can solve displacement fields without training on labeled data, thereby significantly broadening its scope of application. It should be noted that the network architectures for both supervised and unsupervised learning DIC are similar, with speckle images as input and displacement fields as output. This architecture attempts to directly establish a correlation between image pairs and displacement fields, which contracts with traditional Subset-DIC and mesh-DIC methods that employ shape functions to represent the displacement field prior to its optimization. Although neural networks have strong nonlinear fitting capabilities, this "indirect" approach requires neural networks to "extract" information from complex speckle images before "fitting and reconstructing" the displacement field. Consequently, existing supervised and unsupervised learning DIC methods generally require such specially structured neural networks as U-Net [27] and Auto-Encoder [28] to solve the issue. However, studies [24] indicate that solution accuracy remains insufficient even with the deployment of sophisticated networks, necessitating additional displacement continuity constraints to achieve satisfactory results.

Physics-Informed Neural Networks (PINNs) have recently emerged as a more general solution method for systems with governing equations [29]. In PINNs, the coordinate domain of the equation is used as the input to the neural network, with the equation's predicted solution as the output. The physical equations are integrated into the loss function, which steers the neural network through iterative optimization processes until a solution that aligns with the physical laws is achieved. PINNs have demonstrated their efficacy in addressing complex equations across a spectrum of disciplines, encompassing fluid mechanics [30], solid mechanics [31], and inverse problems in mechanics [32-34]. The PINNs approach can be applied to DIC solutions by using the coordinates of the speckle image as the input domain, the displacement field as the output, and the photometric consistency between the two speckle images as the governing equation for constructing the loss function. The DIC solution method based on the PINNs approach, or PINN-DIC, employs a neural network to represent the displacement field and directly solve the displacement field between two speckle images without standard speckle image-deformation field data for training. However, PINN-DIC diverges from unsupervised learning DIC in that it uses coordinates instead of speckle images as input and outputs the displacement field directly, forging an immediate correlation between the displacement field and coordinates. This direct approach confers three advantages. 1) As the network is relieved of the necessity to process intricate speckle images, PINN-DIC can achieve enhanced displacement field representation even with simpler neural network structures, such as fully connected neural networks (FCNN). The uniform approximation property of FCNN for continuous functions ensures a more continuous and smooth representation of the displacement field [35, 36], realizing a more straightforward implementation and higher accuracy. 2) The network architecture that directly inputs the coordinate domain improves the global distribution of points and the imposition of constraints on the coordinate domain. This feature renders PINN-DIC particularly adept at achieving a global, point-by-point solution for irregular boundary speckle images, thereby addressing the limitations of traditional Subset-DIC methods in this regard [16, 37]. 3) The direct input of the coordinate domain simplifies the integration of the DIC solution process with other PINN-based mechanical solutions, such as mechanical simulations and inversion algorithms, facilitating more extensive analysis of DIC measurement data.

In this paper, we develop a DIC solution method using the PINNs approach, presenting the solution concept and algorithm framework. We implement the solution process using Python and the PyTorch library and validate the method's feasibility and accuracy with simulations and real experiments, evaluating its performance through comparisons. The remainder of this paper is structured as follows: The **Method** section introduces the algorithm framework of PINN-DIC and its implementation and validation. The **Assessments** section evaluates the method by comparing it with Subset-DIC and unsupervised learning DIC, assessing PINN-DIC's performance in terms of displacement field accuracy and computational cost. The **Analysis of Deformation Fields for Specimens with Irregular Boundaries** section specifically addresses how PINN-DIC can be adapted for the resolution of displacement fields within speckle images characterized by irregular peripheries. The **Application** section applies PINN-DIC to measure complex deformation fields in real experiments, further demonstrating its effectiveness and advantages. The **Conclusion** provides research findings and future research directions.

## 2 Method

### 2.1 DIC and Its Governing Equations

As shown in Figure 1a, the reference image $I_R(\mathbf{x})$ transforms into the deformed image $I_D(\mathbf{x})$ after undergoing a displacement $\mathbf{u}$. The grayscale values of the two images are in accordance with the following relationship:

$$I_R(\mathbf{x}+\mathbf{u})=I_D(\mathbf{x}) \tag{1}$$

Equation (1) is commonly referred to as the photometric consistency assumption [38]. It implies that the grayscale distribution of spatial points remains unchanged before and after deformation. This assumption holds true in most experimental scenarios, thus forming the governing equation for DIC.

From Equation (1), we can deduce that the fundamental problem of DIC is to solve for $\mathbf{u}$ based on Equation (1) using the $I_R$ and $I_D$ captured from the experiment. It is evident that DIC is inherently an ill-posed problem: for an image of size N pixels × N pixels, as shown in Figure 1b, the number of unknowns in Equation (1) ($2N^2$） exceeds the number of equations ($N^2$). Therefore, additional constraints must be introduced to either increase the number of equations or reduce the number of unknowns to obtain a unique solution. The traditional Subset-DIC method divides the image into many independent subsets and solves the problem for each subset. Given the relatively small size of each subset, it can be assumed that the displacement pattern within each subset is comparatively simple, such as rigid body translation, linear deformation [10], or quadratic deformation [9]. This permits the use of zero-order, first-order, or second-order polynomials (known as shape functions) to describe the displacement of all pixels within a subset with a minimal set of parameters (e.g., 2, 6, or 12). For current Subset-DIC methods, the commonly adopted shape functions are linear displacement, as depicted in the deformed subset in Figure 1b, where the number of unknowns to be solved is significantly smaller than the number of pixels within the subset, rendering the problem solvable.

It is evident from the previous description that the results of the displacement field solution in Subset-DIC are affected by the shape function and subset size. The subset size, particularly, presents a challenge in capturing non-uniform deformations. While larger subsets contain more pixels and offer superior noise suppression, they may cause a mismatch between deformation patterns and the shape function, leading to an over-smoothed displacement field in areas of non-uniform deformation and reduced spatial resolution. Conversely, smaller subsets present a closer match between the deformation and the shape function, but demonstrate limited noise suppression capabilities and lowered resolution in displacement measurement due to their reduced grayscale information. Thus, Subset-DIC is inherently confronted with the challenge of selecting the optimal solution parameters, namely the shape function and subset size, in the context of non-uniform field measurements. In fact, with a fixed shape function and a constant subset size, any selection entails a trade-off between measurement resolution and spatial resolution [39]. In contrast, deep learning-based DIC solutions leverage neural networks to fit the displacement field, effectively employing adaptive shape functions and adaptive subset sizes, thereby enhancing performance in resolving non-uniform displacement fields. [18, 36].

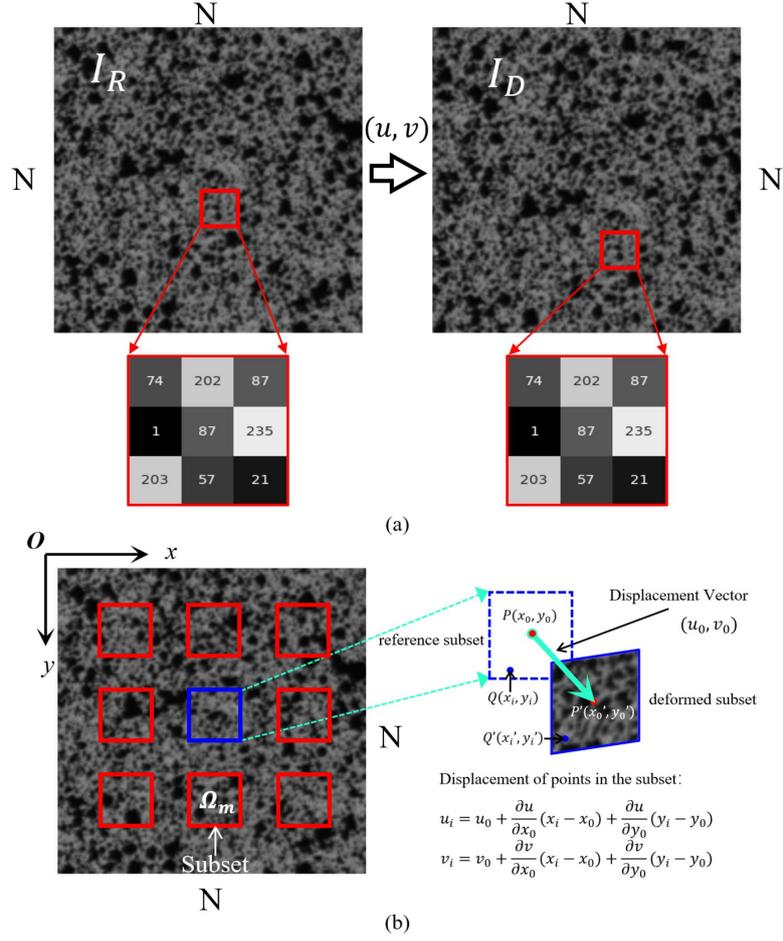

Figure 1. Schematic diagram of DIC principle:

(a) Photometric consistency assumption; (b) Subset-DIC solution approach.

## 2.2 PINN-DIC Framework

The principle of PINN-DIC is shown in Figure 2. The input to the neural network is the coordinate domain **x** that matches the size of the speckle images, and the output is the displacement field **u(x)** of the same size. The loss function $\mathcal{L}$ is defined as:

$$\mathcal{L} = MSE(I_{PD}(\mathbf{x}), I_D(\mathbf{x})) \\
= \frac{1}{L \times H} \sum_{i=1}^{L \times H} (I_{PD}(\mathbf{x}_i) - I_D(\mathbf{x}_i))^2 \tag{2}$$

where $L$ is the width of the image, $H$ is the height of the image, $I_{PD}$ represents the predicted deformed image, $I_D$ represents the actual deformed image, and $MSE$ denotes the mean squared error between the grayscale values of the two images. $I_{PD}$ can be reconstructed from the reference image $I_D$ and the predicted displacement field **u** using an algorithm. A typical reconstruction method can be expressed as:

$$I_{PD}(\mathbf{x}) = \int I_R(\xi)\delta(\xi - (\mathbf{x} - \mathbf{u}(\mathbf{x})))d\xi \tag{3}$$

where $\xi$ is a vector containing the *x* and *y* coordinates of the pixels within the coordinate domain, and $\delta$ represents a Dirac sampling function. Since $\xi = \mathbf{x} - \mathbf{u}(\mathbf{x})$ is usually not at integer pixel locations, it is necessary to interpolate the grayscale values at integer pixel positions near $\xi$ to obtain the grayscale value at the sub-pixel position $\xi$. Common interpolation methods include linear interpolation and cubic interpolation [10].

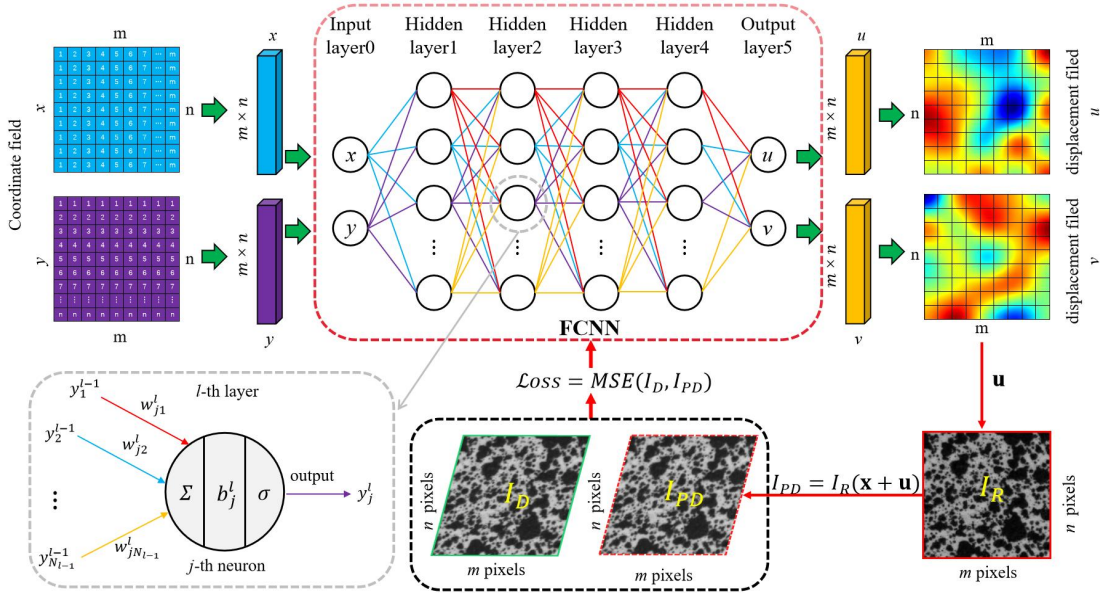

Figure 2. Principle of PINN-DIC.

The PINN-DIC network employs a simple fully connected neural network (FCNN). The network used in this paper consists of 6 neural network layers: an input layer, an output layer, and four hidden layers. The input layer has 2 neurons for inputting the coordinates within coordinate domain, and the output layer also has 2 neurons for outputting the displacements within the coordinate domain. Between the input and output layers are four hidden layers, each with 50 neurons. The output $y_j^l$ of the *j*-th neuron in the *l*-th layer of the FCNN can be expressed as:

$$y_j^l = \sigma\left(\sum_{i=1}^{N_{l-1}} w_{ji}^l y_i^{l-1} + b_j^l\right) \tag{4}$$

where $N_{l-1}$ is the number of neurons in the ($l$-1)-th layer, $y^l_j$ is the output to the $j$-th neuron in the $l$-th layer, $w^l_{ji}$ is the weight of the output of the $i$-th neuron in the ($l$-1)-th layer when computing the $j$-th neuron in the $l$-th layer, $b^l_j$ is the bias of the $j$-th neuron in the $l$-th layer, and σ is the nonlinear activation function, i.e., the adaptive Tanh activation function [40] in this work.

The objective of the PINN-DIC solution is to iteratively optimize the network parameters $\theta(\mathbf{w},\mathbf{b})$ of the FCNN to obtain the optimal displacement field **u** that minimizes the loss value in Equation (2). The solution process is illustrated in Figure 3. First, initial network parameters $\theta_0$ are randomly generated prior to the start of the iteration. At each iteration step, the displacement field $\mathbf{u}_i$ is initialized based on the network parameters $\theta_i$, and the predicted deformed image $I_{PD}$ is obtained using Equation (4). The loss function is then computed, and the network parameters $\theta$ are updated via backpropagation once the loss exceeds a specified threshold. This process is repeated until the loss value or the number of iterations reaches a predetermined threshold.

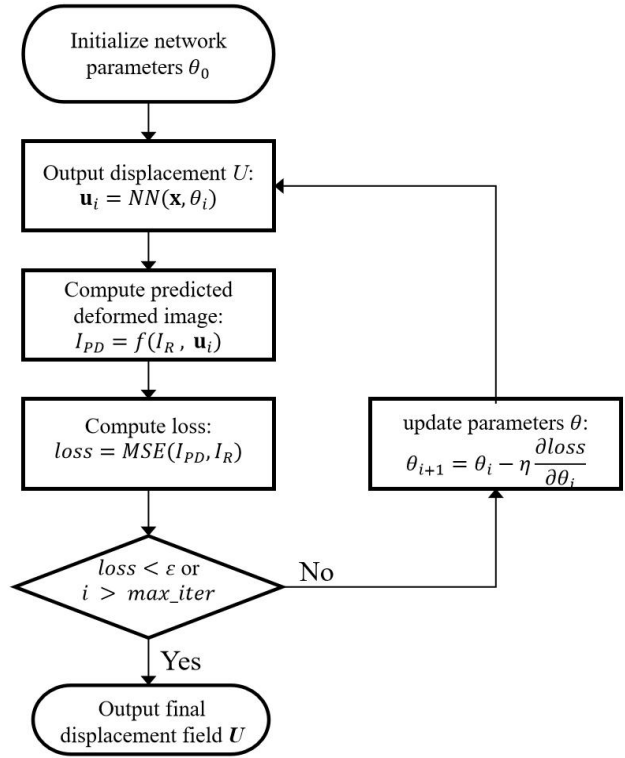

Figure 3. PINN-DIC optimization solution flowchart.

## 2.3 Implementation and Validation of PINN-DIC

The code for implementing the neural network solution described in Section 2.2 was written using PyTorch. During the iterative solution process, two optimizers were employed: the Adam optimizer [41] for initial optimization and the LBFGS optimizer [42] for more refined optimization. This approach effectively enhances convergence speed and improves solution accuracy. For a typical PINN-DIC problem, the specific parameter settings were as follows: the learning rate of the Adam optimizer was set to 1e-3, with a weight decay rate of 1e-4, iterated for 1000 times; and the initial learning rate of the LBFGS optimizer was set to 1, with a maximum of 2000 iterations. To prevent the solution from overfitting to noise, two early stopping conditions were set: 1) the mean gray error between the predicted deformation image and the actual deformation image was less than or equal to 0.1; 2) the range of loss value changes was less than 1% for three consecutive epochs (with each epoch consisting of 100 iterations).

We then used simulated deformed speckle images to verify the feasibility of the PINN-DIC method. The method for generating simulated deformed speckle images is referenced in [19]: 1) A speckle image is generated as a reference image, which can be simulated algorithmically [43] or extracted from real experimental speckle images. 2) A theoretical displacement field was specified, and the grayscale values of the deformed speckle image were interpolated using the method described in Equation (3). 3) Where necessary, noise may be introduced to the interpolated grayscale values to be converted into image grayscale data.

First, we validated the method using simulated deformed speckle images with rigid body displacement. To encapsulate the effects of different speckle characteristics and noise upon the algorithm, six sets of speckle patterns with varying distribution characteristics were harnessed for simulation (see Figures 4a and 4b). Each set of images underwent a rigid body displacement of $u$=0 pixel and $v$=0.2 pixels, and white noise with a mean of 0 and a standard deviation of 2 was added to both the reference and deformed images. The displacement results obtained by PINN-DIC are shown in

Figures 4c and 4d. The results demonstrated the efficacy of this method in addressing different speckle distribution features and noise when solving for rigid body translation displacement fields, achieving high measurement accuracy with absolute errors beneath 0.02 pixels.

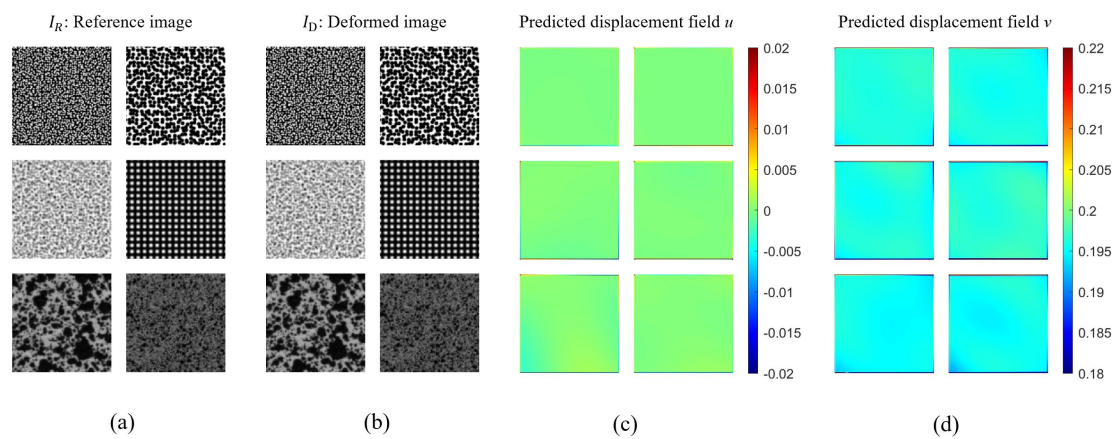

Figure 4. PINN-DIC results for rigid body displacement field analysis: (a)(b) Reference and deformed speckle images with different speckle characteristics (256 pixels × 256 pixels); (c)(d) Solved displacement field $u$ and $v$.

Further validation of PINN-DIC was conducted using speckle images with uniform deformation fields. The simulated deformed speckle images and the applied linear displacement fields are shown in Figure 5a, where both the horizontal displacement $u$ and vertical displacement $v$ vary linearly from -1 pixel to 1 pixel. To evaluate the impact of different noise levels on the algorithm, six noise levels were sequentially applied to both the reference and deformed images (with noise levels of $\mu$ = 0, $\sigma$ = 0,1,2,3,4,5), generating six sets of speckle images. The displacement fields obtained by applying PINN-DIC to these images with varying noise levels are shown in Figure 5b. The results indicated that the method provided high-precision measurements for speckle images across different noise levels and that the displacement field errors increased as the noise level escalated.

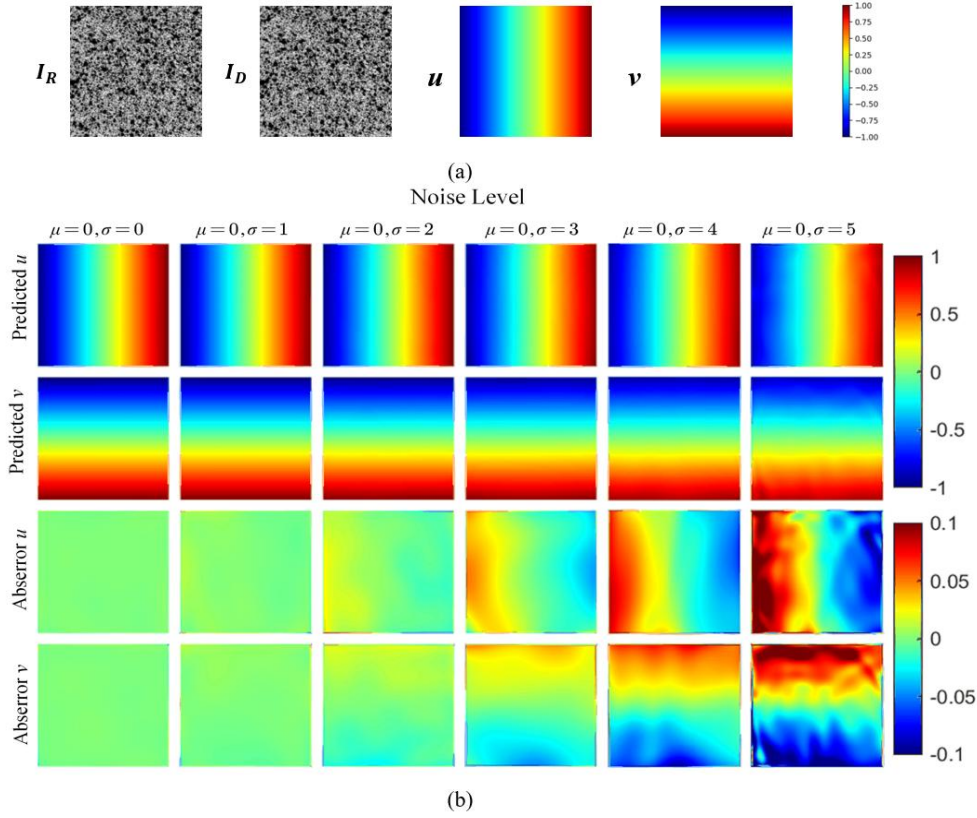

Figure 5. Results of PINN-DIC analysis of uniformly varying displacement fields:

(a) Reference and deformed speckle images (256 pixels × 256 pixels) and applied theoretical displacement field;

(b) Solved displacement fields u and v and their errors.

Subsequently, the capability of the PINN-DIC method to solve non-uniform deformation fields was verified. The speckle images used are shown in Figure 6a, and the applied theoretical displacement field is shown in Figure 6b, with its specific expression as follows:

$$v(x,y) = \cos\left(2\pi \cdot (y - H/2)/A(x)\right) \quad (5)$$

where $H$ is the height of the image,

$$A(x) = P_{min} + x \cdot (P_{max} - P_{min})/L \quad (6)$$

$A(x)$ is the cosine wave's period modulation function, $L$ is the width of the image, and $P_{min}$ and $P_{max}$ are the minimum and maximum periods of the cosine wave in the star-shaped displacement field. In this example, we set $H$=256 pixels, $L$=1024 pixels, $P_{min}$=10 pixels and $P_{max}$=120 pixels. The resulting star-shaped displacement field is shown in Figure 6b. This displacement field exhibited a cosine variation along the $y$-axis, with its frequency attenuating from a higher to a lower magnitude across the

*x*-axis. It can be interpreted that the state of intensity of the non-uniformity with the deformation field decreases from high to low.

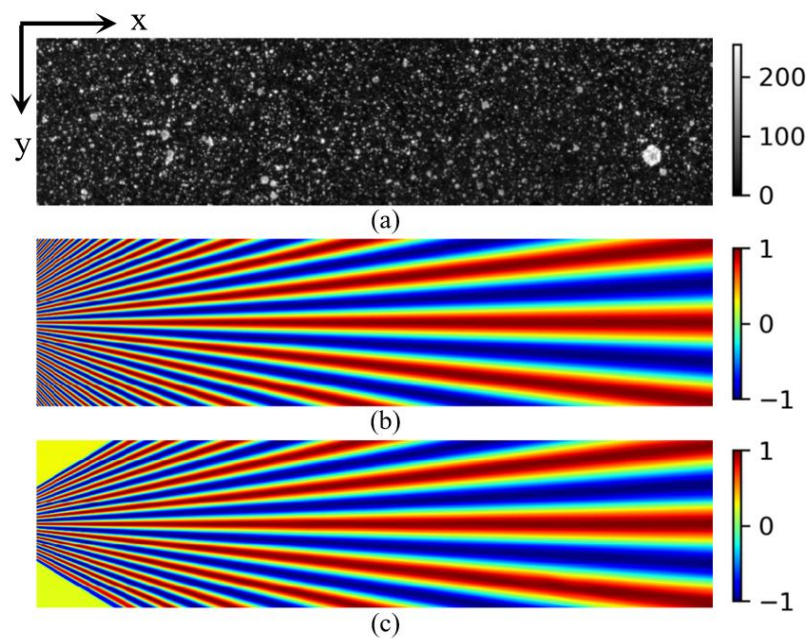

Figure 6. Results of PINN-DIC analysis for non-uniform (star-shaped) displacement field: (a) Reference speckle image (256 pixels×1024 pixels); (b) Theoretical displacement field; (c) Solved displacement field.

When solving the star-shaped displacement field with PINN-DIC, it was observed that the typical settings used previously were ineffective in accurately resolving the high-frequency components of the displacement field, as shown in Figure 6c. To elucidate the underlying cause of this failure, the grayscale residual field **L₁** between the predicted and actual deformed images was plotted at an intermediate iteration phase at the 1200th iteration.

$$\mathbf{L_1} = (\mathbf{I_D}(\mathbf{x}) - \mathbf{I_{PD}}(\mathbf{x})) \odot (\mathbf{I_D}(\mathbf{x}) - \mathbf{I_{PD}}(\mathbf{x})) \tag{7}$$

where $\odot$ is matrix element wise multiplication operator.

As shown in Figure 7a, the residual information presented therein is encapsulated within a histogram (Figure 7b), which exposes a pronounced unevenness in the distribution of the grayscale residual field between $I_{PD}$ and $I_D$, with residual values ranging from $10^0$ to $10^3$. There were fewer points with large residuals on the left side (in the range of $10^2 \sim 10^3$) and lots of points with smaller residuals on the right side. Given that the loss function of PINN-DIC is based on the average grayscale residual, the contribution of the fewer large residual points to the loss can be "diluted" by the whole, thereby leading to a neglect of the left-side regions during the optimization phase. To emphasize the role of the large residual points on the left side in the loss, the loss function needs to be modified. Specifically, the optimization process of PINN-DIC were divided into two stages: the warm-up stage and the formal solving stage. During the initial warm-up phase, the loss function was modified to:

$$\mathcal{L}_1 = \frac{1}{L \times H} \sum_{i=1}^{L \times H} \left( log_{10}(1 + (I_D(\mathbf{x}_i) - I_{PD}(\mathbf{x}_i))^2) \right) \tag{8}$$

where $L$ is the width of the image, and $H$ is the height of the image. By taking the logarithm of the grayscale residuals, the grayscale residual values across the entire field were mapped to a smaller distribution interval, decreasing from a range of $10^0 \sim 10^3$ to an interval of $10^0 \sim 10^1$ (Figure 7c, Figure 7d). This allowed the loss function to equally represent the contributions of points in different regions. In the later formal training stage, the loss function $\mathcal{L}_2$ defined in Equation (2) was used. The maximum number of iterations for both the warm-up stage and the formal solving stage was set to 4000. The thresholds for early stopping of iterations were as follows: iterations were halted when the average grayscale error fell below a threshold of 3 during the warm-up stage and below 0.1 during the formal solving stage. Throughout both the warm-up and formal solving stages, the Adam and L-BFGS optimizers were deployed, while the optimizer parameters remained unchanged from those exemplified in Sections 3.1 and 3.2.

Figure 7f shows the solution results of the PINN-DIC method with the warm-up stage added for speckle images containing a star-shaped displacement field. From the curve of the mean absolute grayscale error decline (Figure 7e), it can be observed that adding the warm-up stage effectively accelerated the convergence rate of the model and improved the solution accuracy.

The results of using the PINN-DIC method with the added warm-up stage to solve the previously mentioned simple deformation fields (rigid body translation and uniform deformation) are essentially the same as those obtained without the warm-up stage. This congruence is rationalized by the fact that both loss functions can equally reflect the contributions of residuals from different regions because simple deformation fields produce a uniformly distributed grayscale residual field with values concentrated within a narrow range during the optimization process. Therefore, the two-stage iterative optimization scheme is used for subsequent solutions.

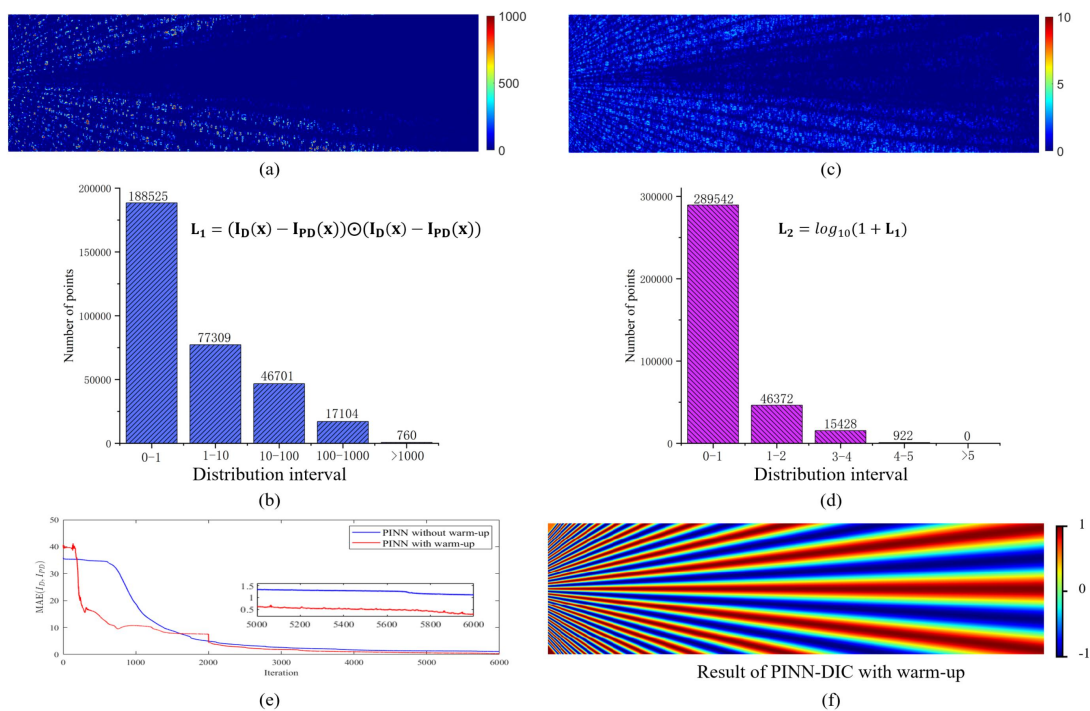

Figure 7. Residuals of the PINN-DIC iterative optimization process: (a) Grayscale residual field of PINN-DIC without the warm-up stage (at iteration 1200) and (b) the corresponding histogram; (c) Logarithm of the residuals in (a) and (d) the corresponding histogram; (e) Average absolute grayscale error reduction curves with and without the warm-up stage; (f) Results of the star-shaped displacement field with the warm-up stage added.

## 3 Assessments

This section evaluates the performance of PINN-DIC through specially designed experiments, including displacement field accuracy and computational costs. To reinforce the credibility of the results, both simulated speckle images with added noise and real speckle images collected from standard experiments are used. PINN-DIC and Subset-DIC are applied to analyze these images, and the results are systematically and quantitatively assessed.

## 3.1 Evaluation Based on Simulated Deformed Speckle Images

Rigid body translation and uniform deformation are relatively simple and have been systematically compared previously. This section focuses on testing the non-uniform (star-shaped) displacement field. The simulated speckle images are shown in Figure 6a, adhering to image size, displacement form, and image generation methods as previously described. To investigate the impact of noise, three levels of noise ($\mu = 0$, $\sigma = 0, 3, 6$) were applied to both $I_R$ and $I_D$, generating three sets of speckle images with different noise levels.

Figure 8 presents the results of PINN-DIC and two sizes of Subset-DIC in analyzing star-shaped displacement fields from noise-free speckle images and those with different noise levels, focusing on the displacement fields and error distributions. It can be observed that: 1) PINN-DIC can resolve displacement fields with diverse degrees of non-uniformity, exhibiting significantly lower errors in displacement field analysis compared with Subset-DIC, notably in regions with higher non-uniformity; 2) As noise levels increase, the displacement field analysis error of PINN-DIC slightly increases, especially in the highly non-uniform regions in the upper left and lower left areas, where Subset-DIC has completely failed; 3) PINN-DIC can resolve displacements throughout the entire region, whereas the presence of sub-regions in Subset-DIC leads to the inability to resolve displacements at the image edges (with a width equal to half the sub-region size). Figure 9 further illustrates the error curves of the displacement fields in three regions with different levels of non-uniformity, corresponding to columns 128, 512, and 896. The analysis reveals that: 1) PINN-DIC demonstrates enhanced noise resistance, evidenced by its smaller errors than those of both sizes of Subset-DIC for all regions and noise levels; 2) PINN-DIC is more suitable for non-uniform field measurements because it exhibits greater difference in errors with Subset-DIC in regions with high displacement field non-uniformity, but smaller difference in more uniform displacement fields.

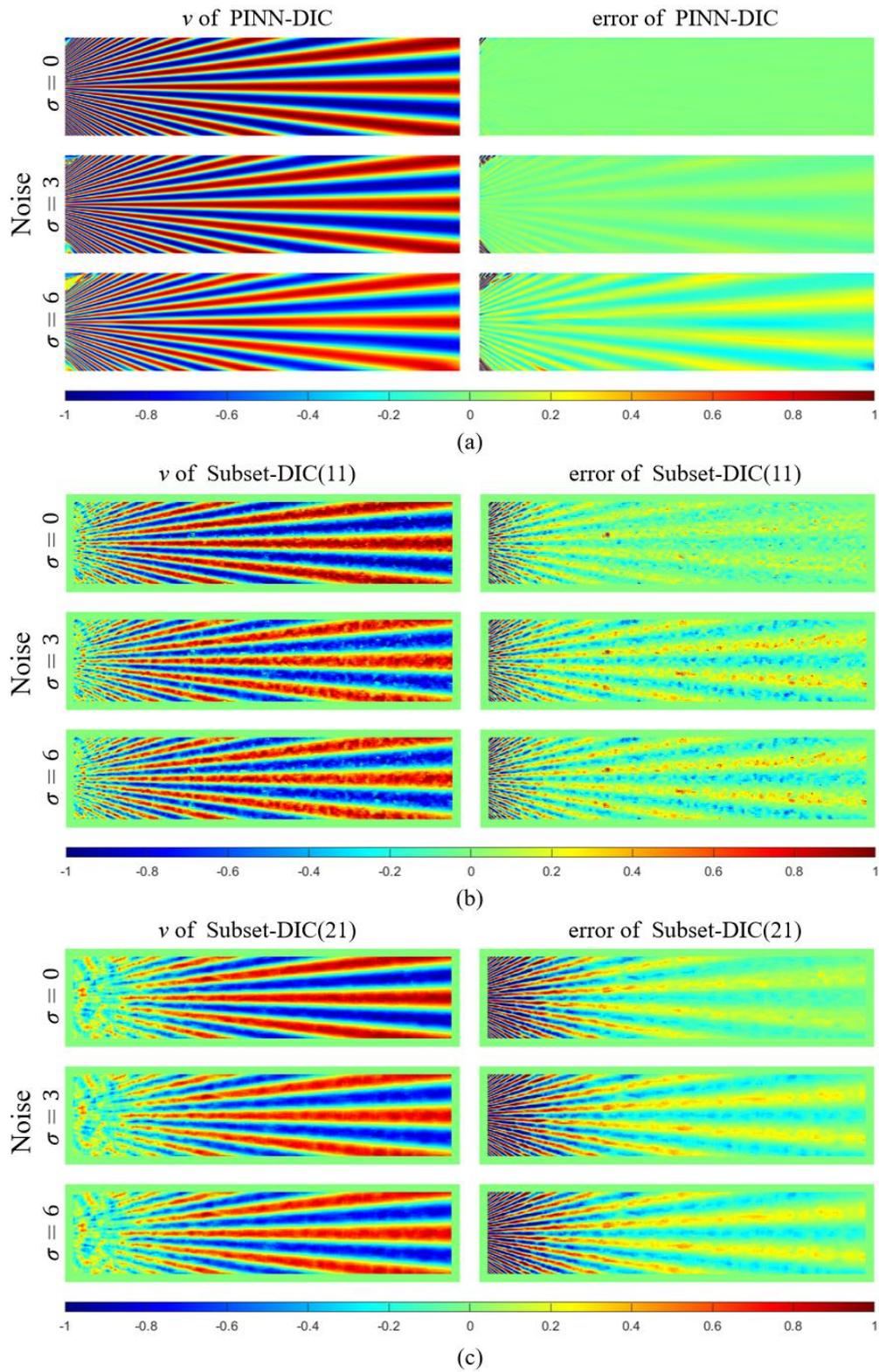

Figure 8. PINN-DIC's capability to analyze non-uniform displacement fields using speckle images with different noise levels (in comparison with Subset-DIC): (a) Results from PINN-DIC; (b) Results from Subset-DIC with a subset size of 11 pixels × 11 pixels; (c) Results from Subset-DIC with a subset size of 21 pixels × 21 pixels.

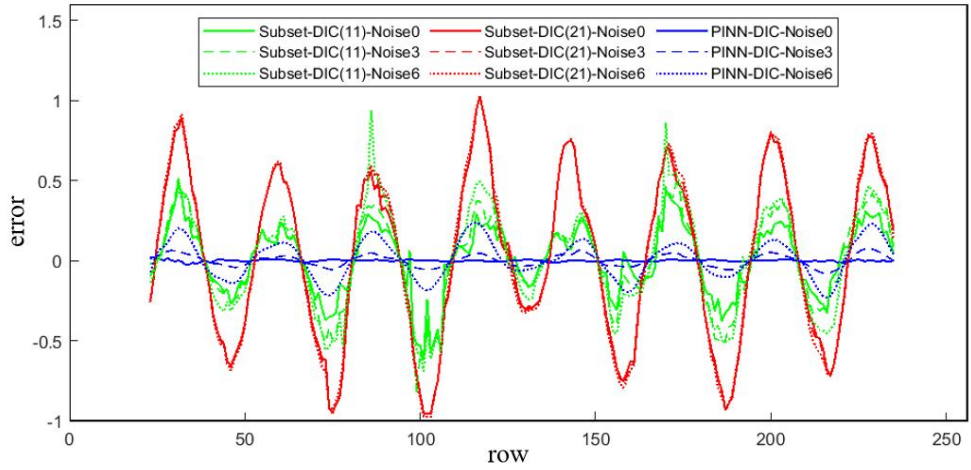

(a)

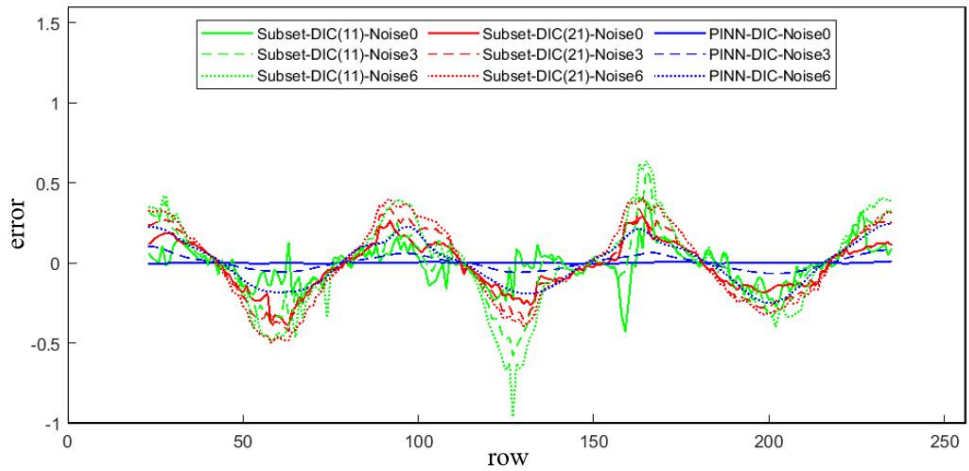

(b)

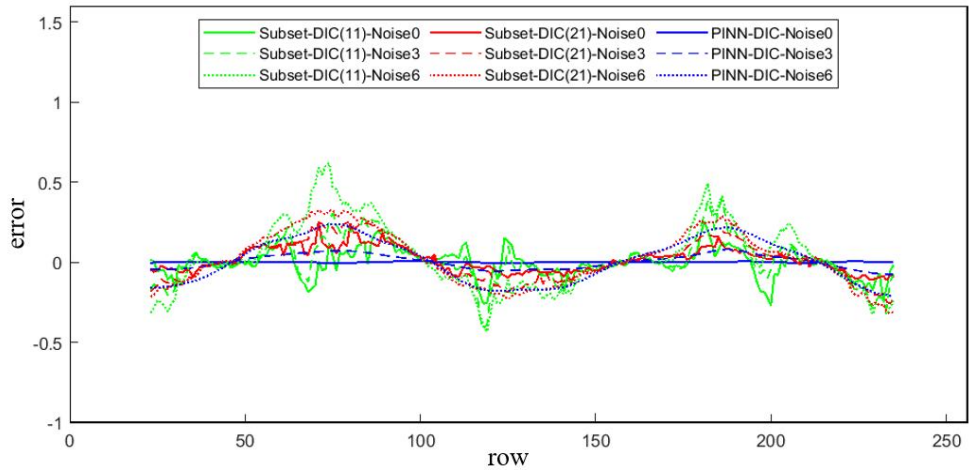

(c)

Figure 9. PINN-DIC's capability to analyze displacement fields with varying degrees of non-uniformity (based on error analysis from different columns in Figure 8): (a) Column 128; (b) Column 512; (c) Column 896.

## 3.2 Evaluation Based on Real Speckle Images

This section evaluates the performance of the PINN-DIC method using real speckle images obtained from a standard four-point bending test. The material used for the four-point bending specimen is 7075 aluminum alloy, with the specimen's structure and dimensions shown in Figure 10a. The specimen is subjected to displacement-controlled loading using a WDW-100 testing machine, with a preload set at 20N and a displacement loading rate of 0.1 mm/min. Images are captured before loading begins as the reference image, and deformed images are captured when the displacement reaches 0.25 mm. A region of interest is selected as shown in Figure 10b, and the resulting speckle images are shown in Figure 10c. The displacement field calculated using PINN-DIC is presented in Figure 11, compared with the displacement fields obtained from finite element analysis and Subset-DIC. The displacement distribution along a selected line is shown in Figure 12.

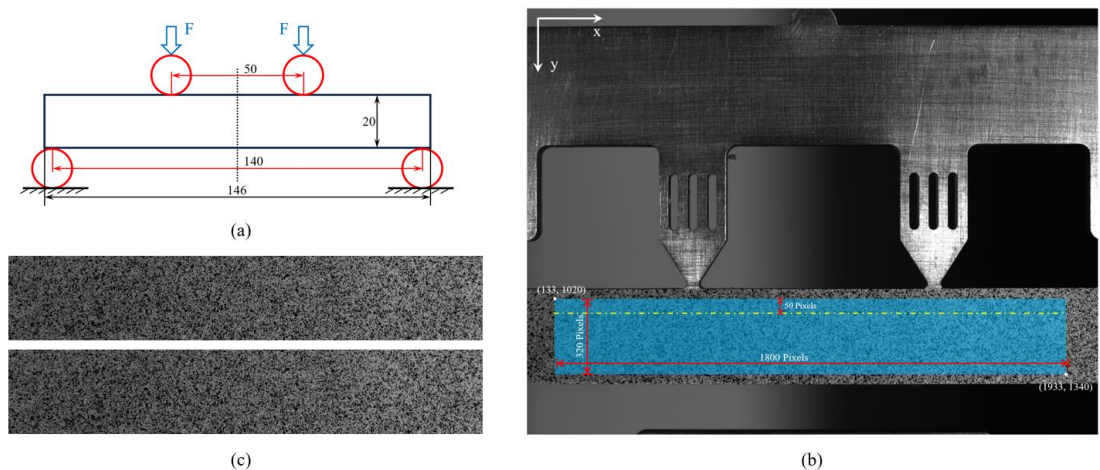

Figure 10. Four-point bending test setup: (a) Specimen and loading schematic; (b) ROI area for DIC analysis;

(c) Reference and deformed speckle images.

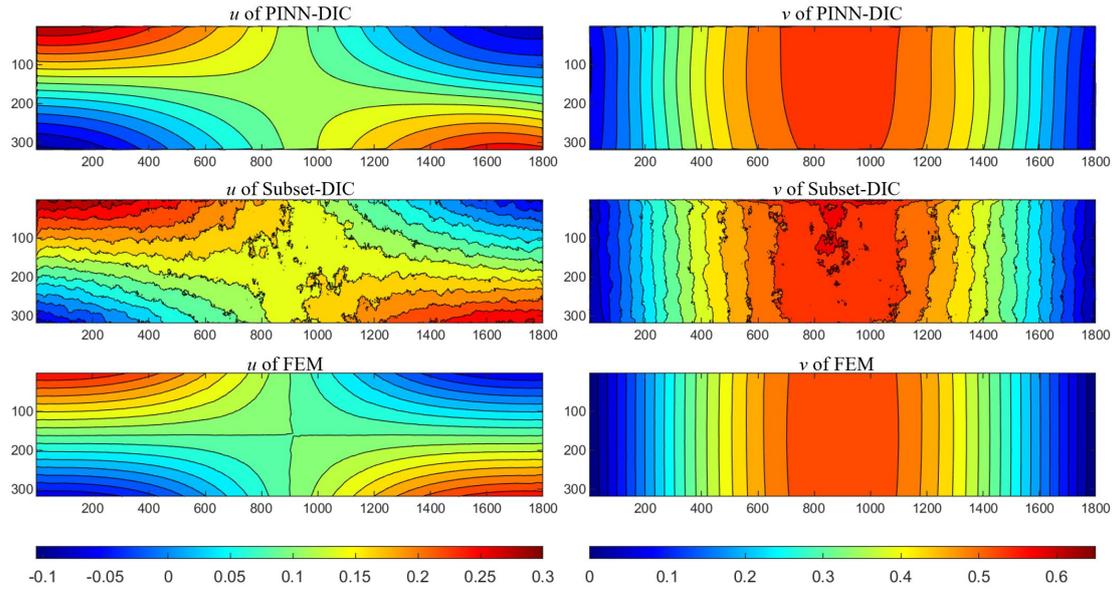

Figure 11. Displacement field analysis results: the first row are PINN-DIC results; the second row are Subset-DIC results (subset size: 31 pixels × 31 pixels); the third row are Finite element simulation results (for comparison).

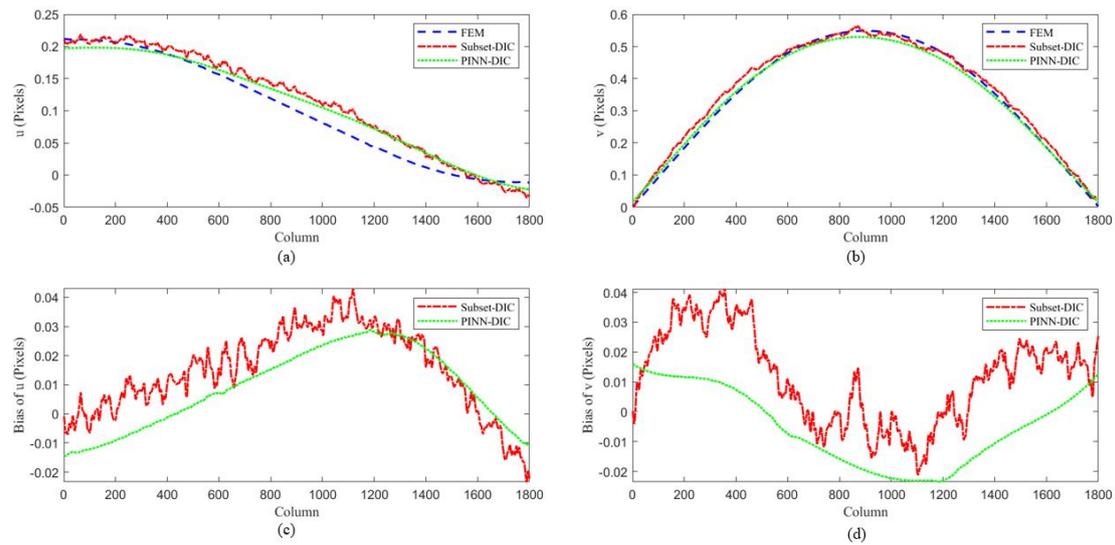

Figure 12. Comparison of displacement results along a selected section (dashed line in Figure 10(b)): (a) Displacement $u$; (b) Displacement $v$; (c) Error in displacement $u$; (d) Error in displacement $v$.

From Figures 11 and 12, it is evident that the displacement analysis results obtained using PINN-DIC are smoother compared with those from Subset-DIC. Furthermore, the deviations of the PINN-DIC results from the theoretical values provided by FEM results are smaller than those from Subset-DIC. This experiment demonstrates that PINN-DIC maintains strong applicability when processing real speckle images, delivering more favorable outcomes in comparison to Subset-DIC.

## 3.3 Computational Cost Analysis

This section evaluates the computational cost of the PINN-DIC method in analyzing displacement fields. The test is conducted by comparing the computational cost of PINN-DIC with that of traditional Subset-DIC methods, represented by Ncorr [44] and Vic-2D [45], using the same speckle images on a computer equipped with an NVIDIA GeForce RTX 3050 6GB GPU. To ensure the reproducibility of the experiment, the solution process for the PINN-DIC method in all cases involves the following: the Adam and L-BFGS optimizers are each used for 2000 iterations during the warm-up phase and reapplied during the formal training phase with learning rates and threshold parameters detailed in Section 2.3. The key parameters for solving with the Subset-DIC method are set as follows: the subset size is 31 pixels × 31 pixels and the step size is 1. The computational cost is evaluated by the processing speed, defined as the average number of points processed per second (Points/s). Considering the variance in the ROI sizes of PINN-DIC and Subset-DIC, the processing speed for Subset-DIC is calculated excluding the boundary points, as illustrated in Figure 13.

Figure 14 compares the processing speeds of different methods when handling four different types of displacement fields. It reveals that the processing speed of PINN-DIC is comparable to that of Subset-DIC for the latter three types of deformation, much slower than the commercial software only for the task of rigid body deformation, which is rarely encountered in practical measurement. It is acknowledged that this comparative analysis is somewhat rudimentary, as it does not meticulously account for the influence of iteration termination conditions on processing speed. Nonetheless, this elementary assessment still underscores the practicality of PINN-DIC, which secures high-precision measurement outcomes without incurring an increased computational cost in comparison with Subset-DIC.

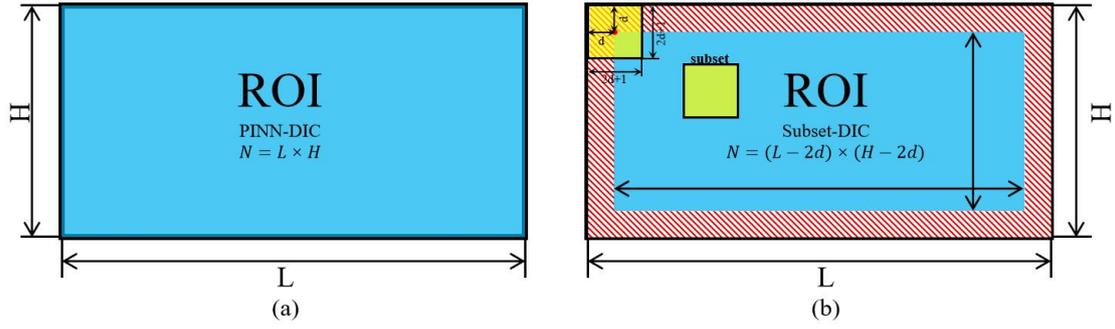

Figure 13. Schematic of Calculation Regions: (a) ROI for PINN-DIC; (b) ROI for Subset-DIC.

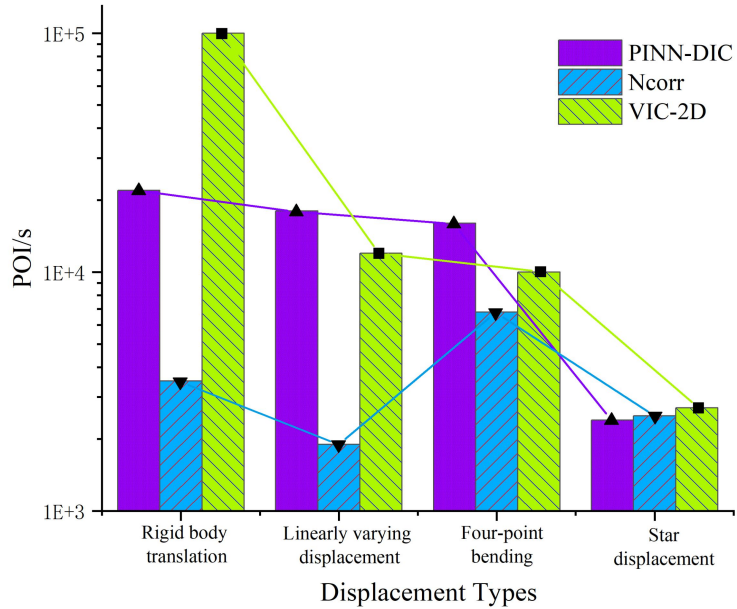

Figure 14. Comparison of computational costs.

## 3.4 Comparison with Unsupervised Learning DIC

The differences between Unsupervised Learning DIC (referred to as USL-DIC) and the PINN-DIC method described in this paper are predominantly attributable to their respective network architectures, as illustrated in Figure 15. In USL-DIC, the speckle images serve as the input and the displacement field is the output, while PINN-DIC inputs the coordinate domain and outputs the displacement field. Consequently, the network in USL-DIC needs to extract features from the speckle images before fitting the displacement field, whereas the network in PINN-DIC directly fits the displacement field. Drawing upon relevant theoretical research on neural networks, the latter network structure is more streamlined with superior performance in fitting complex continuous nonlinear functions [46]. This section uses the unsupervised learning DIC network architecture from Cheng et al.'s research [24] (code available at: https://github.com/cxn304/USL_DIC) as a reference to compare the differences in implementation and performance between PINN-DIC and unsupervised learning DIC.

Using the fully connected network architecture described in Section 2.2 of this paper and the U-net network in the original text of Cheng et al., unsupervised learning DIC (USL-DIC) programs were developed to solve the speckle images with the star-shaped displacement field described in Section 3.1 (as shown in Figure 16). The analysis of the results, when compared with those obtained using the method outlined in this paper, reveals the following: 1) When using the same simple network as PINN-DIC, USL-DIC is nearly incapable of producing satisfactory results; 2) Even with the more complex U-net network, the displacement fields computed by USL-DIC exhibit lower accuracy and smoothness and heightened vulnerability to noise than those produced by the PINN-DIC method. A review of the programming implementations reveals that the fully connected network architecture used in this paper is considerably less complex than the U-net network employed by Cheng et al., with respective parameter counts of 5,356 and 31,044,034, a difference of nearly 6,000 times.

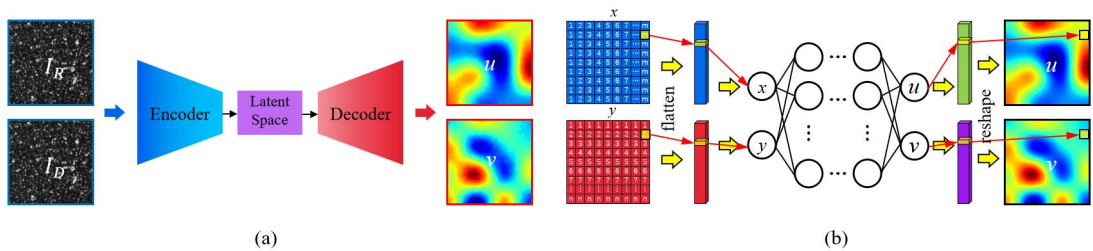

Figure 15. Differences between the two deep learning DIC network architectures: (a) USL-DIC network based on CNN encoder-decoder; (b) PINN-DIC network based on FCNN.

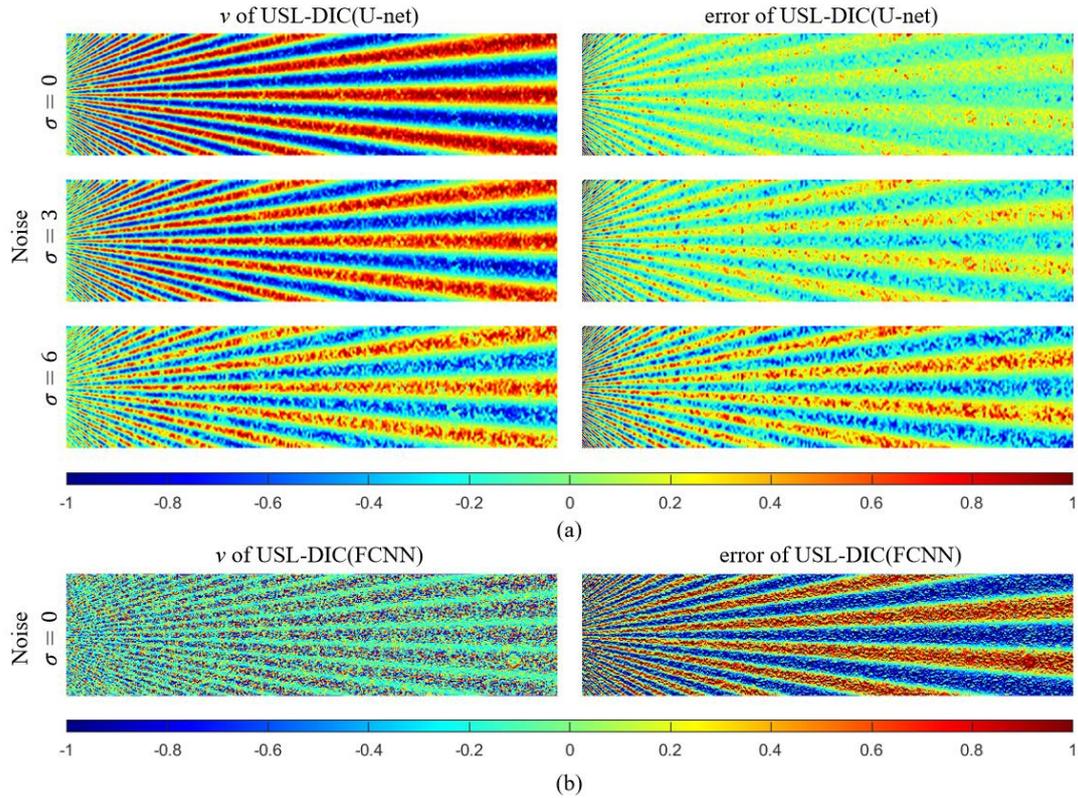

Figure 16. Results of USL-DIC for the star-shaped displacement field:

(a) Using U-net network; (b) Using FCNN network.

# 4 Analysis of Deformation Fields for Specimens with Irregular Boundaries

Measuring deformation fields for specimens with irregular boundaries is essential in mechanical analysis but has posed a significant challenge for traditional Subset-DIC methods [4, 16, 37]. This challenge arises because subsets frequently extend beyond the image limits. Traditional Subset-DIC typically employs two strategies to address this issue: 1) avoiding the boundaries by processing only those regions where subsets do not exceed the image boundary [45], which implies that displacements at the boundaries remain unmeasured; 2) "shrinking" the subsets by using smaller subsets or partial subsets at the boundaries to estimate displacements [47]. However, the accuracy of displacement estimation is compromised or even rendered unfeasible because the "amount of information" within subsets decreases at the boundaries. Supervised learning methods for estimating deformation fields from images with irregular boundaries [23] require training with images of that are identical in shape and size, thereby limiting their practical applicability. In contrast, the PINN-DIC framework, which has the coordinate domain as its input, can effortlessly manage images with irregular boundaries, as boundary constraints can be incorporated directly into the input coordinate domain by removing coordinates outside the boundary from the domain matrix.

The specific operation of using PINN-DIC to analyze deformation fields from images with irregular boundaries is illustrated in Figure 17. The process involves manually or automatically identifying and marking the specimen's boundary, inputting the coordinates of the image within the boundary into the network, and then conducting the displacement field analysis following the same PINN-DIC analysis procedure as above. Figure 18 shows experimental results of using PINN-DIC to analyze the deformation field of a circular ring specimen under radial compression, in comparison with results from two Subset-DIC boundary handling methods. The figure demonstrates that the PINN-DIC method can accurately determine the displacement field across the entire domain, including specimens with irregular boundaries. Conversely, the two Subset-DIC methods either fail to determine the displacement at boundary points or present less accurate displacement measurements in these areas.

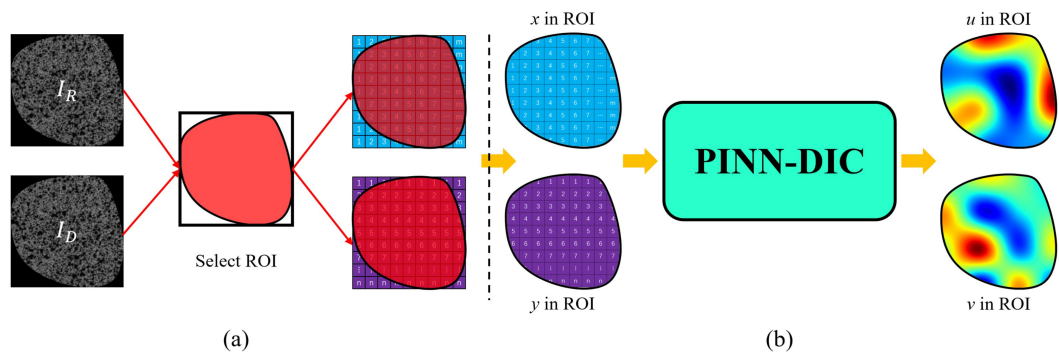

Figure 17. Illustration of PINN-DIC for analyzing deformation fields with irregular boundaries: (a) Irregular boundary identification; (b) PINN-DIC for irregular boundary.

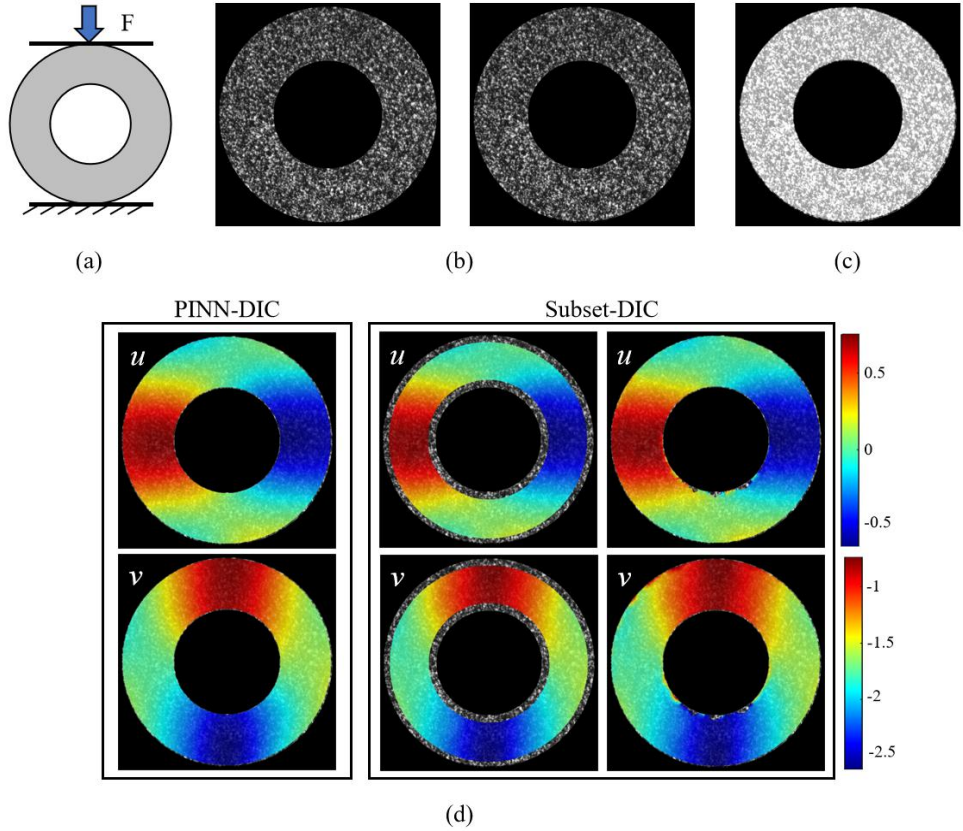

Figure 18. Displacement field measurement experiment for radially compressed circular ring specimen: (a) Loading schematic of the specimen; (b) Deformed and reference speckle images; (c) Computational region; (d) Comparison of analysis results.

# 5 Application

As established in the preceding sections, PINN-DIC presents considerable advantages over traditional Subset-DIC, particularly in the measurement of specimens with irregular boundaries and non-uniform deformation. Building on the verification of these advantages, this section demonstrates the application of PINN-DIC to measure the deformation field evolution throughout the loading process in line contact structures.

As shown in Figure 19, a cylindrical specimen and a cubic specimen, both fabricated from polyoxymethylene (POM) material, are in line contact to form a typical test specimen for a line contact structure, with speckles applied to their end surfaces. The test specimen is loaded using the WDW-100 testing machine in displacement control mode, with a loading rate of 0.2 mm/min. During the experimental loading process, the end surface of the contact structure test specimen is imaged at 1 fps by an IPX-16M3-L digital camera with a resolution of 4872 pixels × 3248 pixels, until noticeable plastic deformation is observed. After the experiment, the reference image and deformation images under typical loads are selected (Figures 19c and 19d), key processing areas are identified (Figure 19e), and the displacement field is obtained using PINN-DIC. This displacement field is then used to calculate the strain field and analyze the yield process within the contact structure.

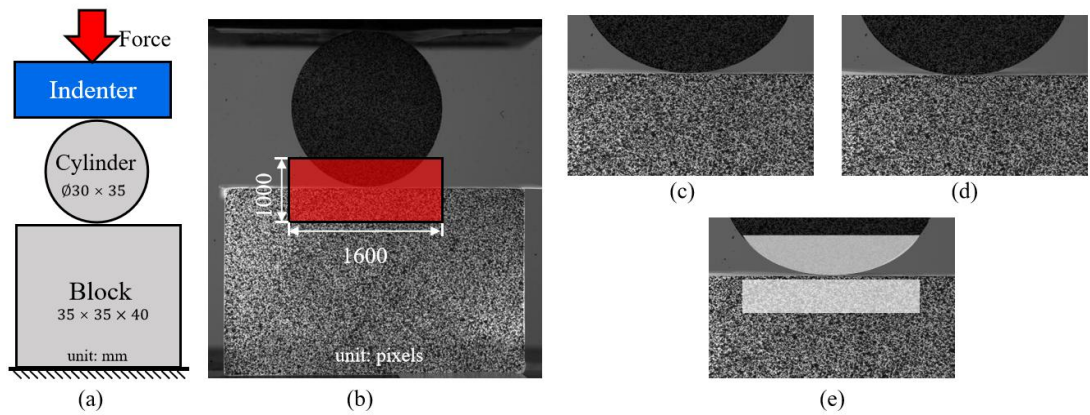

Figure 19. Line contact experiment setup: (a) Schematic of specimen model loading; (b) Camera-captured images and analysis region; (c) Reference image; (d) Deformation image; (e) Selected ROI.

Figure 20 shows the displacement field of the line contact structure at a typical moment (point C in Figure 21a). The observation reveals that PINN-DIC can accurately determine the displacement field across the complex boundary region of the contact specimen's end face in a single attempt, concurrently delivering high-resolution measurements within regions of non-uniform deformation (near the contact zone).

Figure 21 shows the evolution of the maximum shear strain field at five typical loading stages. The figure reveals the nucleation and propagation of the yield zone in the two parts of the line contact structure, including the upper cylindrical section and the bottom cubic section. Based on the experimental results, an in-depth analysis of the yield process of the contact structure can be conducted.

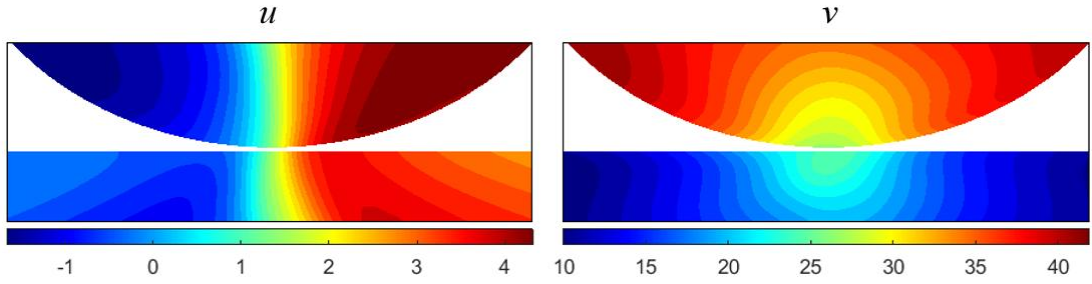

Figure 20. Displacement field measurement results.

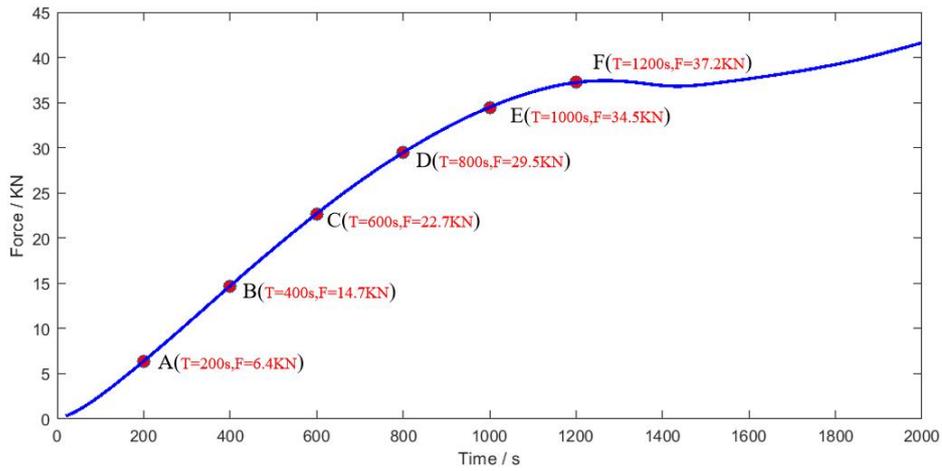

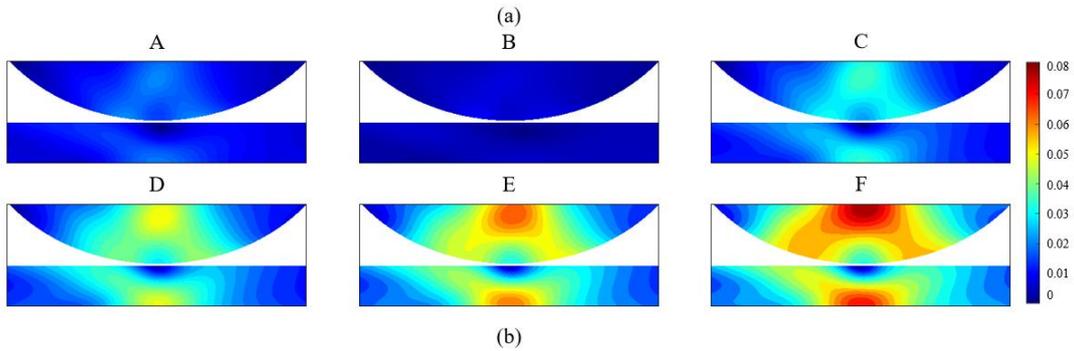

Figure 21. Evolution of the maximum shear strain field: (a) Force loading curve; (b) Maximum shear strain field distribution at six moments.

Within the observation area of a line contact structure characterized by irregular boundaries and severe non-uniform deformation, traditional subset-DIC method processes the two parts separately, demonstrating low accuracy in the displacement field analysis. In this experiment, the PINN-DIC approach was deployed to process the speckle images of the line contact structure, enabling the simultaneous acquisition of a high-precision displacement field for the entire irregular region. This highlights the distinct advantage of PINN-DIC.

# Conclusion

In this study, the Physics-Informed Neural Network (PINN) method is introduced to the Digital Image Correlation (DIC) approach, resulting in the development of PINN-DIC. This approach integrates the photometric consistency assumption of speckle images as a governing equation within the neural network framework through a loss function, enabling PINN-DIC to directly solve the displacement field from reference and deformed speckle images. This paper constructs the framework of PINN-DIC, implements the solving process through programming, and develops a two-stage optimization method based on two different loss functions, thereby enhancing PINN-DIC's adaptability in addressing non-uniform displacement fields. The feasibility of PINN-DIC is verified using both simulated images with various displacement fields and noise levels and real images with different speckle characteristics. The evaluation of the accuracy and computational efficiency of the method indicates that compared with Subset-DIC, PINN-DIC effectively resolves the conflict between spatial resolution and measurement resolution, achieving more favorable results when processing non-uniform deformation fields. Furthermore, PINN-DIC exhibits better noise resistance than Subset-DIC. In terms of computational expense, PINN-DIC's cost is comparable to that of Subset-DIC; while it is slightly slower in complex displacement field calculations, its speed is either equivalent to or surpasses that of Subset-DIC in simple displacement field calculations.

The article conducts a comparative analysis of the network architecture and solving approach of PINN-DIC in relation to the unsupervised learning DIC (USL-DIC) method. The comparison reveals that PINN-DIC attains greater accuracy with a more streamlined network structure. This advantage stems from the fact that USL-DIC takes speckle images as input and outputs the displacement field, requiring a complex network architecture for feature extraction from the images, which in turn escalates the model's complexity and the quantity of parameters. In comparison, PINN-DIC directly uses the coordinate domain as input and outputs the displacement field, thereby obviating the requirement for image feature extraction and simplifying the network architecture. Furthermore, the fully connected neural network architecture of PINN-DIC presents higher accuracy in fitting complex, smooth, and continuous nonlinear functions, thereby generating smoother and more continuous calculation results. Even when USL-DIC employs a more complex U-net network, its accuracy and smoothness in solving the displacement field remain inferior to those of PINN-DIC and is more susceptible to noise. Therefore, PINN-DIC demonstrates superiority over USL-DIC in terms of computational complexity, network simplification, and solving accuracy, rendering it a more effective method for solving DIC problems.

Due to the network architecture's direct input of the solution domain, PINN-DIC can solve speckle images with irregular boundaries through a straightforward adjustment of the input data, i.e., selecting coordinates located within the ROI region. This article compares the performance of PINN-DIC and Subset-DIC in calculating the displacement field of a circular ring specimen. The results show that PINN-DIC can achieve stable and accurate solutions even at the boundaries, demonstrating its significant advantages in handling complex geometric boundaries. Finally, PINN-DIC is applied to a practical line contact yield experiment with POM material specimens, verifying its effectiveness in real experimental environments. Furthermore, PINN-DIC exhibits remarkable performance in measuring the non-uniform deformation field of specimens with irregular boundaries, proving its advantages in complex stress analysis.

PINN-DIC introduces popular techniques from the AI for Science field into DIC solving. It not only provides DIC with a higher precision and broader applicability approach but also establishes a standardized framework for DIC. Such a framework can be conveniently connected with other methods in the field, such as inverse problem solving and numerical-experimental integration. This development is anticipated to foster the evolution and broader application of DIC methodologies. Nonetheless, this paper represents an initial attempt into the application of PINN to DIC, with the full potential of PINN not yet being fully harnessed in this work. Future research could explore the incorporation of additional constraints of deformation field or high-precision measurement results into the PINN framework to further optimize DIC solutions. Moreover, there is potential to develop an integrated inverse problem-solving framework grounded in PINN, which could directly solve structural mechanical parameters from speckle images, thereby broadening the analytical capacity of DIC measurement outcomes.

## Acknowledgments

This research was partially supported by National Natural Science Foundation of China (Grant Nos. 12132009, 12272046), and the Deep Blue Project Foundation of Shanghai Jiao Tong University of China (Grant Nos. SL2022ZD101).